\documentclass[aps,prd,twocolumn,showpacs,preprintnumbers,superscriptaddress,nofootinbib]{revtex4}
\usepackage{amssymb}
\usepackage{bm}
\usepackage{enumerate}
\usepackage{amssymb}
\usepackage{amsmath}
\usepackage{feynmf}
\usepackage{slashed}
\usepackage{wrapfig}
\usepackage{graphicx}
\usepackage{color}
\newcommand\be{\begin{equation}}
\newcommand\ee{\end{equation}}

\newcommand\bea{\begin{eqnarray}}
\newcommand\eea{\end{eqnarray}}

\begin{document}

\title{CDM and Baryons as Distinct Fluids in a Linear Approximation for the Growth of Structure}

\author{Soma De}
\email{somad@asu.edu}
\affiliation{Department of Physics \& School of Earth and Space Exploration, Arizona State University, Tempe, AZ 85287-1404}

\author{James B.\ Dent}
\email{jbdent@louisiana.edu}
\affiliation{Department of Physics \& School of Earth and Space Exploration, Arizona State University, Tempe, AZ 85287-1404}
\affiliation{Department of Physics, University of Louisiana at Lafayette, Lafayette, LA 70504-4210, USA}

\author{Lawrence M. Krauss}
\email{krauss@asu.edu}
\affiliation{Department of Physics \& School of Earth and Space Exploration, Arizona State University, Tempe, AZ 85287-1404}

\begin{abstract}

A single fluid approximation which treats perturbations in baryons and dark matter as equal has sometimes been used to calculate the growth of linear matter density perturbations in the Universe.  We demonstrate that properly accounting for the separate growth of baryon and dark matter fluctuations can change some predictions of structure formation in the linear domain in a way that can alter conclusions about the consistency between predictions and observations for $\Lambda$CDM models versus modified gravity scenarios .   Our results may also be useful for 21cm tomography constraints on alternative cosmological models for the formation of large scale structure.

\end{abstract}



\maketitle 

\section{Introduction}

Since the discovery that the universe is apparently dark energy-dominated causing an observed acceleration (\cite{Perlmutter:1998np,Riess:1998cb}, also inferred indirectly on the basis of other observational constraints, i.e. see, for example \cite{Krauss:1995yb}), a vast expenditure of effort has been made towards possible explanations of the acceleration.  The standard paradigm of cold dark matter with a cosmological constant ($\Lambda$CDM) in the cosmological framework of general relativity (GR) accommodates all experimental evidence, and remains the simplest and most economical cosmological model consistent with the data.  Though the $\Lambda$CDM cosmology fits all the present data, issues such as the hierarchy and coincidence problems remain which highlight the issue of how the acceleration can be realized in a fully consistent theoretical framework.  This has led to a consideration of alternative cosmological models, including models in which gravity varies away from GR on large scales. As observations become increasingly precise, the $\Lambda$CDM picture will be put through ever more rigorous tests in the effort to constrain new physics.  It is important therefore to have accurate theoretical frameworks by which to judge whether observations may indicate a discrepancy with the predictions of the standard model.  

By now a standard way to constrain various cosmological alternatives is via an exploration of the growth of linear matter perturbations for various redshifts.   These perturbations have been parameterized via a growth index (see for example \cite{Peebles:1980aa,Wang:1998gt,Linder:2007hg,Bertschinger:2006aw,Polarski:2007rr,Dent:2009wi}), which has a specific value for $\Lambda$CDM  (to first order in deviations from a purely CDM dominated universe, where the deviation is due to a cosmological constant, this index has been estimated to be simply 6/11).

The growth is typically found in the following manner.  One defines a matter overdensity given in $k$-space by $\delta \equiv \delta\rho/\rho$ with $\rho$ being the background matter density.  Using the standard Einstein equations, one finds the dynamical equation typically called the growth equation, for a single component matter field in a matter dominated universe
\begin{eqnarray}\label{growth1}
\ddot{\delta} + 2H\dot{\delta} -4\pi G_N\rho\delta = 0
\end{eqnarray}
where the overdot is a derivative with respect to coordinate time and we
have dropped the $k$ index.   This relation is given in the synchronous
gauge where it holds on all scales (for gauge related issues see for
example
\cite{Dent:2008ia,Yoo:2009au,Chisari:2011iq,Bonvin:2011bg,Challinor:2011bk,Jeong:2011as,Yoo:2010ni,Perivolaropoulos:2010dk}).
From here one can define a function $g\equiv {\rm{d ln}}\delta/{\rm{d ln}} a \equiv \Omega_m(a)^{\gamma}$ which leads to the equation
\begin{eqnarray}\label{growth2}
g' + g^2 + g\left(\frac{\dot{H}}{H^2} + 2\right) = \frac{3}{2}\Omega_m
\end{eqnarray}
where the prime denotes a derivate with respect to the natural log of
the scale factor. The function $g$ can be identified as the growth
factor for matter density perturbations. The solution to this for a flat universe with a dark energy equation of state $w_{\Lambda}$ is given, to first order in the expansion parameter $1-\Omega_{\Lambda}$ by 
\begin{eqnarray}
\gamma = \frac{3(w_{\Lambda}-1)}{6w_{\Lambda}-5}
\end{eqnarray}
This relation reduces to  $\gamma = 6/11$ for the case of $\Lambda$CDM.
The growth factor $g$ defined in (2) is affected $only$ by CDM overdensities. In our paper we will consider the effect of considering both
CDM and baryonic perturbations on the growth factor.

Therefore we focus on two facts:

\begin{itemize}

\item The matter content of the universe is not solely composed of cold dark matter, as the baryonic content is roughly one-fifth that of dark matter \cite{Komatsu:2010fb}.  If one writes separate growth equations for dark matter and baryonic matter densities,  each contain a source term involving the gravitational potential which is a function of the full matter content, including both baryons and dark matter (for example, the third term in Eq.(\ref{growth1}) arises from the Poisson equation for the gravitational potential).  

\item  Because  baryonic matter has a non-zero sound speed, dark matter and baryonic matter perturbations obey different dynamical equations \cite{Mukhanov:1990me,Ma:1995ey,Naoz:2005pd}.

\end{itemize}

Here we explore the consequences of properly incorporating both
of these effects, and quantify and compare differences in the
perturbative densities when the full set of baryonic plus dark matter
equations are solved versus the case when baryons are ignored.   One of
the central purposes of this effort is to compare the relative differences
between results of these two approaches compared to the differences obtained
when different cosmological models are explored, in order to determine the sensitivity to cosmological model dependence versus the need to properly account for baryons.

To explore these effects we perform calculations under both the standard $\Lambda$CDM scenario
and a modified gravity model, for which we choose the DGP  \cite{DGP2000} model. We compare bias factors, total matter
density perturbations and the growth factor in these to cosmological scenarios
scenarios to explore the sensitivity to not including baryonic perturbations. 
We find that an accurate treatment of baryonic
fluctuations will alter quantities like the bias factor and the total
matter density fluctuation even in the linear regime in a way that can exceed the change induced in the quantities by varying the the background
cosmology.  On the other hand we find that the growth factor, defined as
$\frac{dln\delta}{dlna}$ is relatively insensitive to inclusion or non-inclusion of baryonic dynamics over linear scales. It is on the other hand very
sensitive to the background cosmological model.

In Section II we briefly present the formalism for the full set of
dynamical equations.  In Section III we present our results, and finally in Sec. IV we
conclude with a brief discussion of their implications.

\section{Calculation}
The full set of coupled linear differential equations for the growth of
perturbations in dark matter ($\delta_c$) and baryons ($\delta_b$) along
with radiation ($\delta_{rad}$) is
\begin{eqnarray}
\label{e:cosmo1}
\ddot{\delta_{c}}+2H(z)\dot{\delta_{c}}&=&\frac{3}{2}H^{2}(f_{c}\delta_{c}+f_{b}\delta_{b}+f_{r}\delta_{rad}) \nonumber \\
\ddot{\delta_{b}}+2H(z)\dot{\delta_{b}}+c_{s}^{2}k^{2}&=&\frac{3}{2}H^{2}(f_{c}\delta_{c}+f_{b}\delta_{b}+f_{r}\delta_{rad}) \nonumber \\
\delta_{rad}&=&\frac{4}{3}\delta_{b} 
\end{eqnarray}
\begin{eqnarray}
\label{e:cosmo2}
f_{c}&=&\Omega_{c}(z) \ \ \ \ f_{b}=\Omega_{b}(z) \ \ \nonumber\\ 
f_{r}&=&\Omega_{rad}(z),0 \ (z>z_{dec},z<z_{dec})
\end{eqnarray}
where $\Omega_{c}$, $\Omega_{b}$ and $\Omega_{rad}$ are CDM, baryons and
radiation energy density at a given epoch. We define $c_{s}=\left(\frac{\delta P}{\delta\rho}\right)_{S}$
\cite{Ma:1995ey}, where the subscript $S$ stands indicates that $c_{s}$ is defined 
at constant entropy, $S$. At low redshifts
$c_{s}$ is primarily due to baryonic pressure and at
high-redshift(pre-decoupling redshifts), the contribution is mainly from radiation pressure. 
At high redshifts we use
Eqns.(\ref{e:cosmo1}) and (\ref{e:cosmo2}) to solve for the baryonic and dark matter
perturbations. We evolve the perturbations in each Fourier mode $k$, starting
from the epoch of horizon entry for that particular mode.
Therefore our boundary values for $\delta_{b}$ and
$\delta_{c}$, corresponding to a mode $k$, are set at the epoch of
horizon entry, $z_{ent}$, such that 
\begin{eqnarray}
\label{e:cosmo3}
A&=&3\delta_{COBE} \nonumber\\
k_0&=&\frac{a(z_{eq})H(z_{ent})}{c} \nonumber\\
\delta_{c,b}(z_{ent},k)&=& A\left(\frac{k}{k_0}\right)^{\frac{n_{s}}{2}}
\left(k<k_{eq}\right)\nonumber \\
&=& A\left(\frac{k}{k_0}\right)^{\frac{n_{s}-4}{2}} \left(k>k_{eq}\right)\nonumber \\
\dot{\delta_{c,b}}(z_{ent},k)&=& \delta_{c,b}(z_{ent},k)H(z_{ent})
\left(z_{ent}>z_{eq}\right)\\\nonumber
&=&\delta_{c,b}(z_{ent},k)a(z_{ent})H(z_{ent}) \left(z_{ent}<z_{eq}\right)
\end{eqnarray}
In Eq.(\ref{e:cosmo3}), modes that enters the horizon at the epoch of
matter and radiation equality ($z_{eq}$) are denoted by
$k_{0}$. Fluctuations are normalized by using the temperature
fluctuation over angular scales of 7 degrees at the surface of last
scattering measured by the COBE mission which is denoted by $\delta_{COBE}$ \cite{COBE1990}. The spectral index of the primordial fluctuations (coming from
very early times) is set by $n_{s}$ over all Fourier modes. 

The boundary conditions for $\delta_{c,b}$ described by Eq.(\ref{e:cosmo3})
 come from the following argument. Given a primordial power spectra
of shape $P_{i}(k) \propto k^{n_{s}}$, fluctuations grow during the radiation
dominated epoch such that $\delta \propto a^{2}$. Therefore when a
given mode enters the horizon, its power is described by $P_{ent}(k)
\propto a^{2}P_{i}(k) \sim k^{n_{s}-4}$ if $z_{ent} > z_{eq}$
\cite{Peebles1993}, and we replace
$\delta_{c,b}(z_{ent},k) \propto \sqrt{P_{ent}(k)}$. Note that if $n_s \approx 1$, then  $k^{3}P_{ent}(k)$ is constant at horizon crossing.  Similarly, we
find that, $P_{ent}(k) \propto k^{n_s}$ corresponding to the modes
which enter the horizon after matter-radiation equality.

The calculation of $c_{s}$ involves matter and radiation temperatures along
with their fluctuations.
We calculate the matter temperature, $T_{mat}(z)$, at a given epoch
$z$ using the following equation \cite{Seager2000}.
\begin{eqnarray}
\label{e:cosmo4}
\frac{d\bar{T}_{mat}}{dt}&=&2H(z)\bar{T}_{mat} \nonumber \\
&+&\frac{x_{e}(t)}{t_{\gamma}}\left(
\bar{T}_{\gamma}-\bar{T}_{mat}\right)a^{-4} \\
\end{eqnarray}
The radiation temperature, $T(z)$, at a given epoch $z$ is estimated by the
standard relation $T(z)=T_{0}(1+z)$, with $T_{0}$ given by the 
present CMB temperature \cite{Bennett2012}.
Fluctuations in the matter temperature after mechanical decoupling
($z>1100$), $\delta_{T}$, are calculated using
\begin{eqnarray}
\frac{d\delta_{T}}{dt} &=&
\frac{2}{3}\frac{d\delta_{b}}{dt}-\frac{x_{e}(t)}{t_{\gamma}}a^{-4}\frac{\bar{T}_{\gamma}}{\bar{T}}\delta_{T} 
\end{eqnarray}
$\delta_{T}$ thus computed is in turn used for the calculation of the sound
speed post recombination, and therefore the modified baryonic
growth equation becomes \cite{Naoz2005}
\begin{eqnarray}
\label{e:cosmo6}
\ddot{\delta_{b}}+2H(z)\dot{\delta_{b}}&=&
\frac{3}{2}H^{2}(f_{c}\delta_{c}+f_{b}\delta_{b}) \nonumber\\
&-& \frac{k^{2}}{a^2}\frac{k_{B}T}{\mu}(\delta_{b}+\delta_{T})
\end{eqnarray}
For scales which enter the horizon before the epoch of recombination,
$z_{rec}$, matter temperature fluctuations are described by $\delta_{T}=\delta_{T_{\gamma}}$ at $z=z_{ent}$.

As we would like to examine growth not only in the standard cosmology, but in 
a modified gravity scenario as well, we will now discuss how the calculation needs to be altered.
In the DGP scenario, the CDM and baryon perturbation equations are modified
such that in the source term on the right hand side of Eq.(\ref{e:cosmo1}),
the factor $H(z)^{2}$ is replaced by $H^{2}_{DGP}g(a,k)$. In this
case $H_{DGP}(z)$ is the modified background expansion rate and $g(a,k)$
is the factor by which Newton's constant gets modified under the new gravity
scenario. 

We use \cite{Hu2007} and \cite{HuSong2007} 
to construct $H_{DGP}(z)$ and $g(a,k)$. Using these modifications and
Eqns.(\ref{e:cosmo1}-\ref{e:cosmo2}), we calculate the growth of perturbations in the
DGP theory up to a scale corresponding to $k<0.05$Mpc$^{-1}$. We chose to
restrict ourselves to these scales in order to avoid complications due
to non-linear PPF parameters as described in \cite{Hu2007}. 
The matter temperature perturbations in Eqns.(\ref{e:cosmo4}-\ref{e:cosmo6})
also get modified by replacing $H(z)$ with
$H_{DGP}(z)$.

The input parameters of our calculation are $n_{s}$, $\delta_{T_{\gamma}}$, 
$\Omega_{b,c}$, $h$,
$\Omega_{k}$, $\Omega_{tot}$ and $z_{eq}$. 
Additionally, the epoch of decoupling, $z_{dec}$, is determined such that the photon mean-free path is
larger than Hubble distance, or
$\lambda_{\gamma}=\frac{1}{n_{e}\sigma_{T}} \sim cH^{-1}$. 
For simplicity, we set $\Omega_{\nu}=0$,
$\frac{dn_{s}}{dlnk}=0$ and allow a sharp drop in optical depth of
photons at the epoch of mechanical decoupling. In the next section we
will describe our results using WMAP9 values \cite{Bennett2012}, along with additional DGP fits.

\section{Results} 

\begin{figure}[htbp]
\centering{\includegraphics[width=3.0in]{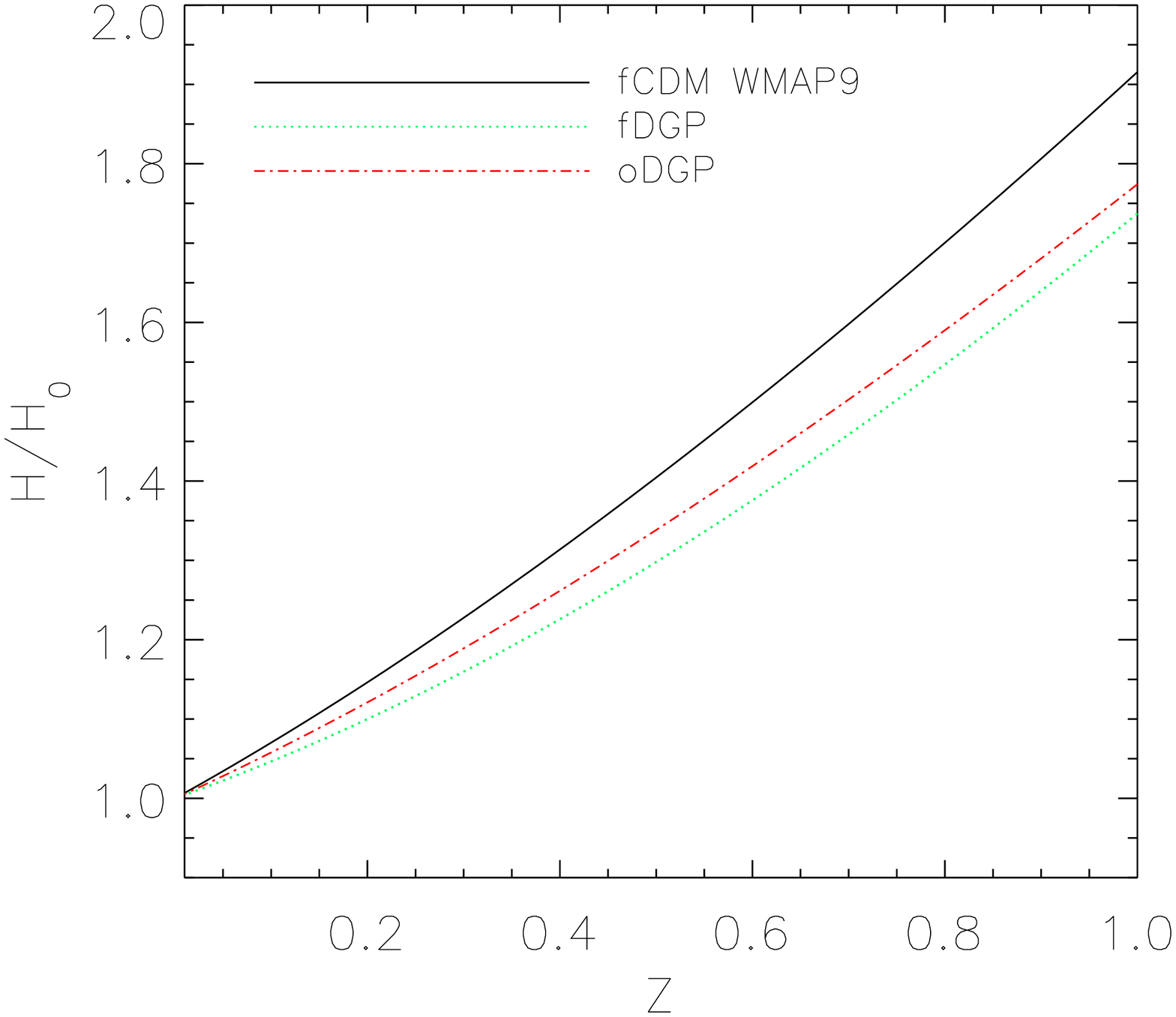}}
\caption{Background expansion rates
($H/H_{0}$) as a function of redshift based on different fCDM (flat
$\Lambda$CDM), fDGP (flat DGP) and oDGP(open DGP) cosmologies. Relevant
cosmological parameters are described in the text.
}
\label{plot1}
\end{figure}

In Figure \ref{plot1} we represent the background expansion rate with respect to
redshift, $H(z)$,
in different cosmological cases. We consider fCDM, fDGP
(flat $\Lambda$CDM and flat DGP), and oDGP (open DGP) models. For cosmological
parameters we use
$\Omega_{c}h^{2}=0.12$, $\Omega_{b}h^{2}=0.023$, $h=0.69$ from WMAP9, and
apply those to both fCDM and fDGP models. For the oDGP models we use
$\Omega_{c}h^{2}=0.099$, $\Omega_{b}h^{2}=0.023$, $\Omega_{k}=0.03$  and
$h=0.76$ \cite{HuSong2007}, and for the fDGP
model we use $\Omega_{c}h^{2}=0.12$,$\Omega_{b}h^{2}=0.023$,
$h=0.69$.  Eqn.(4-5) of \cite{HuSong2007} are incorporated to compute the modified
expansion rate $H_{DGP}(z)$.
We have plotted up to $z\sim 1$ to highlight
the effect at low redshift, where one would expect modifications of gravity designed to
mimic dark energy to be most relevant. 

\begin{figure}[htbp]
\centering{\includegraphics[width=3.0in]{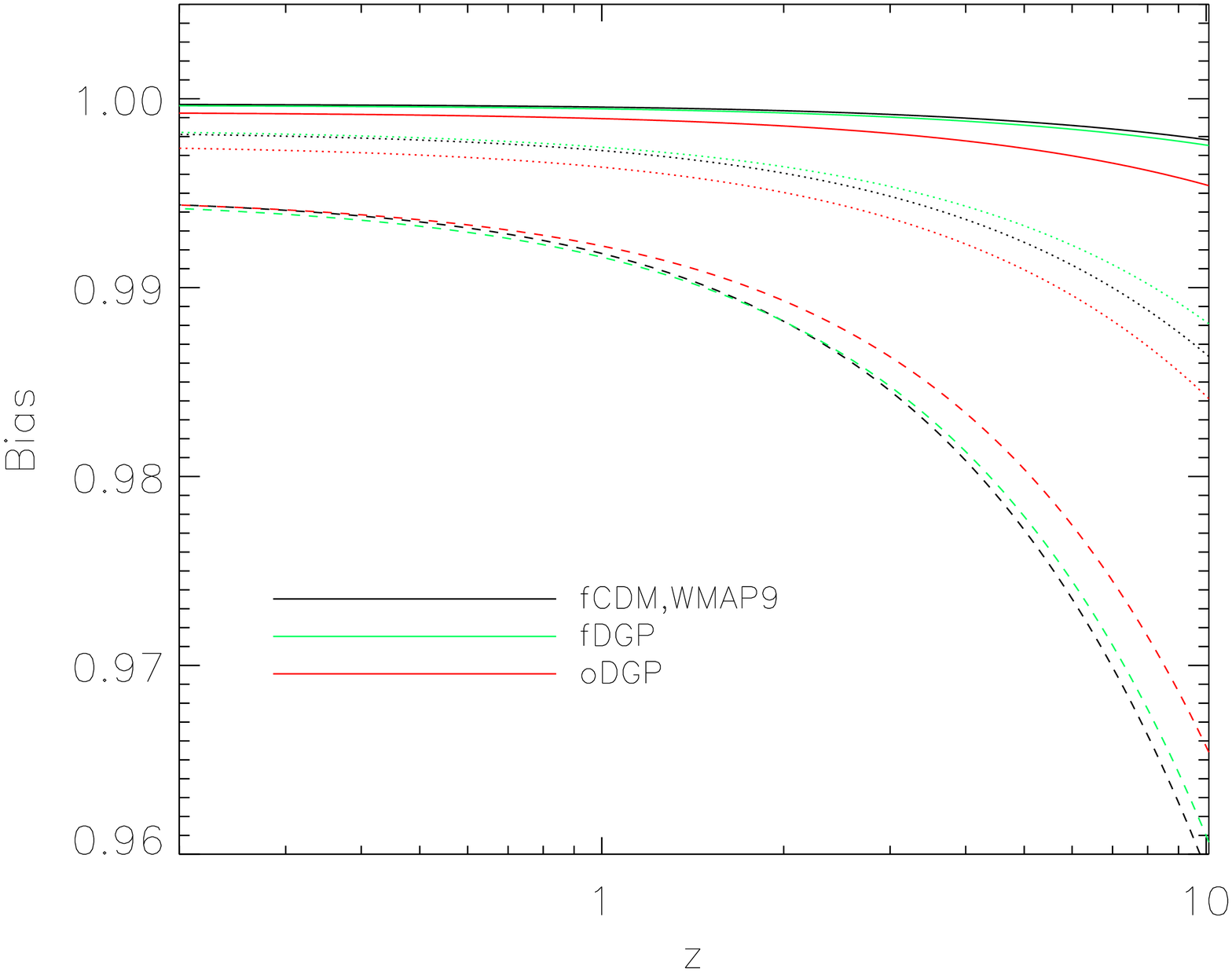}}
\caption{The bias as a function of redshift in different cosmologies for scales corresponding to
$k=0.005$(solid lines), $k=0.01$(dotted lines) and $k=0.05$(dashed
lines) in units of Mpc$^{-1}$.}
\label{plot2}
\end{figure}

In Figure \ref{plot2} we present the bias, defined as
$b(z,k)=\frac{\delta_{b}}{\delta_{c}}$, with respect to redshift. We choose
three length scales corresponding to Fourier modes $k=0.005$, $k=0.01$ and $k=0.05$ in
units of Mpc$^{-1}$, where $b(z,k)$ is represented in those regimes respectively by solid,
dotted and dashed lines. We choose fCDM,
fDGP and oDGP cosmologies described by the same cosmological parameters
as in Figure \ref{plot1}. 
For a single fluid model the bias is unity by construction. From Figure \ref{plot2} it is evident that
in all scales explored, the difference in bias between the  DGP and CDM
models is less than their deviation from unity. This deviation (with
respect to
unit bias in a single-fluid model) increases both with redshift and diminishing scale. 
For low redshift ($z \sim 0.2$) and $k=0.01$Mpc$^{-1}$, the difference between the bias
calculated from fCDM and fDGP models (using WMAP9 parameters) is about
$0.04 \%$ 
with a
difference of $4.2 \%$ in the background expansion rate. The bias becomes close to
1\% near $z \sim 1$, while the difference in bias between fCDM and fDGP
(using WMAP9 parameters) is only up to 0.02\% for
$k=0.05$Mpc$^{-1}$.

\begin{figure}[htbp]
\centering{\includegraphics[width=3.0in]{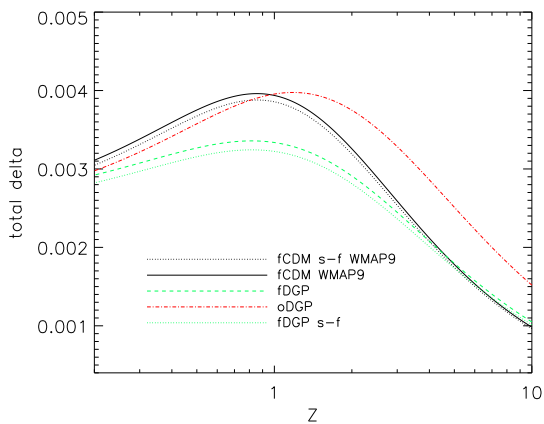}}
\caption{The total matter fluctuation 
$\delta=f_{c}\delta_{c}+f_{b}\delta_{b}$ is shown as a function of redshift under different cosmologies at a scale corresponding to
$k=0.005$Mpc$^{-1}$. Solid lines indicate
two-fluid models and dotted lines represent single fluid (s-f) models.}
\label{plot3}
\end{figure}

The bias can be directly related to the total density fluctuation
$\delta=f_{c}\delta_{c}+f_{b}\delta_{b}$ such that
$\delta=\left( f_{c}+b(z,k)f_{b} \right)\delta_{c}$, where $b(z,k)$ represents
the bias at a given epoch and a given scale $k$. In Figure
\ref{plot3} we display $\delta_{tot}(z,k)$ as a function of
redshift in the linear regime, $k=0.005$Mpc$^{-1}$. The same 
normalization was
used for all cosmologies, set by Eq.(\ref{e:cosmo3}). We use
dotted lines to refer to the single fluid (s-f) models and solid lines
for the baryon+CDM fluid models. We note that at low redshift, the
difference in the oDGP and fCDM models is comparable to that between the 
fCDM, s-f and two fluid models. This is an intriguing conclusion which suggests 
the importance of the two-fluid treatment in order to correctly use structure formation observations to constrain cosmological models. The significance of $\delta_{tot}$ is that
it is a scale and cosmology dependent quantity. Observationally, future
weak lensing surveys can estimate $\delta_{tot}$ but the interpretation of
observational data must be made by properly incorporating bias on all
scales of interest.

\begin{figure}[htbp]
\centering{\includegraphics[width=3.0in]{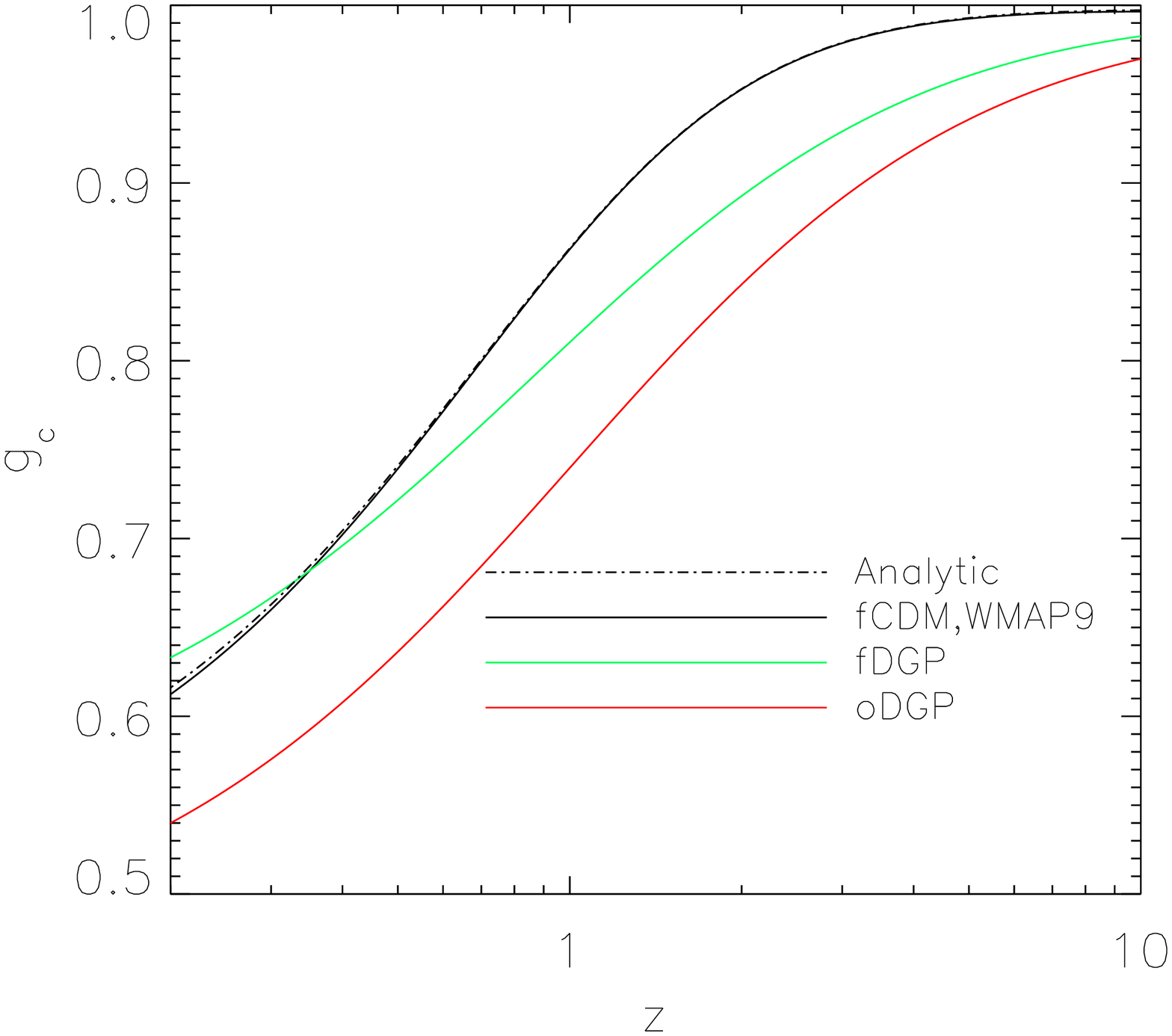}}
\caption{The growth factor due to $only$ CDM described by $g_{c}=\frac{dlnl\delta_{c}}{dlna}$ is plotted as a function of redshift for different fCDM (flat
$\Lambda$CDM), fDGP (flat DGP) and oDGP(open DGP) cosmologies.  This is done at a single scale $k=0.005$Mpc$^{-1}$.Two
component models are presented in solid lines and the dash-dotted line
referred as analytic indicates $\Omega_{m}(z)^{\gamma}$ such that $\gamma=\frac{6}{11}$ for fCDM using
WMAP9 parameters.
}
\label{plot4}
\end{figure}

In Figure \ref{plot4}, we plot growth factor $only$ due to CDM $g_{c}=\frac{dln\delta_{c}}{dlna}$ (solid
lines) with respect to redshift for a chosen scale corresponding to
$k=0.005$Mpc$^{-1}$. Note that we over plot (dashed line) the parameterization
$\Omega_{m}^{\frac{6}{11}}$ \cite{Peebles1980} which agrees quite well with the
numerical estimates for fCDM. We therefore conclude
that $g_{c}$ is roughly scale-independent, but is sensitive to the background cosmology. 

We can also consider the evolution of the total growth factor due to CDM
and baryons as $g_{tot}=\frac{dln\delta_{tot}}{dlna}$ with respect to redshift at a
scale of $k=0.005$Mpc$^{-1}$, to check to see if there is any difference in using this value instead of the CDM growth factor.  We define
$\delta_{tot}=\frac{\delta_{c}f_{c}+\delta_{b}f_{b}}{f_{c}+f_{m}}$.  We find that both
parameters $g_{c}$ and $g_{tot}$ are weakly dependent on the selection of
scale $k$., with $g_{tot}$ to be comparatively more rigid over a range of various
$k$ values.  However in general $g_{c} \sim
g_{tot}$ so there is no significant handicap in using $g_c$ to constain cosmological models.

\section{Discussion and Future Directions}

In this work we have examined the impact on calculations of the growth of structure through the
use of a simple single matter fluid approximation vs. a model that correctly incorporates
baryons and cold dark matter in a two component analysis.  We have then compared the 
difference in matter perturbations in the standard
$\Lambda$CDM cosmology and modified DGP gravity.

Among the quantities we discussed in our paper, the growth factor
$g=\frac{dln\delta}{dlna}$ is measured observationally from 
galaxy redshift surveys \cite{Colless2001}. We found the growth factor,
$g$ to be rigid with variation in $k$ under both $\Lambda$CDM and DGP
cosmologies, and largely independent of baryonic dynamics.

 Using weak lensing to measure the total
$\delta=f_{c}\delta_{c}+f_{b}\delta_{b}$  is a way to get a handle
on total matter density fluctuations but this quantity is sensitive to baryonic dynamics
(\ref{plot3}) and depends on the relevant scale of structure formation
$even$ $in$ $the$ $linear$ $regime$. In addition, we find that at low-redshift ($0.7 < z < 1.5$) the
effect of baryonic dynamics can be comparable to that introduced by
modifying the underlying cosmology. We show in Figure \ref{plot3} that
for $z \sim 1$, modifications due to two component model is comparable
to that due to modifying the cosmology. Therefore accurate
inclusion of the baryonic dynamics is required for interpreting observations
associated with this quantity

It is also worth noting various studies of the growth of large scale structures in different 
modified gravity contexts have been performed using a single growth equation 
(for a sample see 
\cite{Gong:2008fh,Gannouji:2008wt,Fu:2009nr,Wu:2009zy,Motohashi:2010qj,
Gupta:2011kw,BuenoSanchez:2010wd,DiPorto:2011jr,Koyama:2005kd,Zhang:2011di}).
 Preliminary results, to be described in a future work suggest that the use of a proper two component fluid formalism can significantly weaken the ability to distinguish between cosmological predictions in such models.

Finally, another area of cosmology which has attracted a great deal of interest 
recently, and relies on accurate calculations of the evolution of density 
perturbations, is the signal arising from the 21cm spin-flip transition of 
neutral hydrogen (for recent reviews, see \cite{Pritchard:2011xb,Furlanetto:2006jb,
Morales:2009gs}).  One may study, for example, the perturbations of the 
brightness temperature of the CMB over a large redshift range in the so-called 
Dark Ages, which rely on the density perturbations of hydrogen.  These 
perturbations are seeded by dark matter, and therefore a precise 
calculation merits the inclusion of the full baryonic plus dark matter system.  We are currently using the formalism we have described here to investigate how it will impact upon conclusions one may draw from the use of such observations.

\bigskip
We thank S. White for useful conversations.  J. D and LMK were supported in part by a DOE grant to ASU. S.D. was supported in part by funds from SESE Explorer Fellowship.
\bibliography{growth}
%
%
%
%
%
%
%
%
%
%
%
%
%
%
%
%
%
%
%
%

\end{document}